\documentclass[fleqn,twoside,twocolumn]{elsart}
\usepackage{graphicx}
\usepackage{epsfig}
\DeclareSymbolFont{greek}{U}{eur}{m}{n}
\DeclareMathSymbol{\upalpha}{\mathord}{greek}{"0B}
\DeclareMathSymbol{\upbeta}{\mathord}{greek}{"0C}
\DeclareMathSymbol{\upgamma}{\mathord}{greek}{"0D}
\DeclareMathSymbol{\updelta}{\mathord}{greek}{"0E}
\DeclareMathSymbol{\upmu}{\mathord}{greek}{"16}
\newcommand{\um}{$\upmu$m} 
\begin{document}
\begin{frontmatter}

\vspace*{-11mm}{\it \begin{flushleft} \small
Talk presented at the 6th International Conference on Large Scale Applications
and Radiation Hardness of Semiconductor Detectors, 
September 29-October 1 2003, Firenze, Italy. 
\end{flushleft}
}\vspace*{-6.5mm}

\title{Tests of silicon sensors for the CMS pixel detector}
\author[Zurich,PSI]{A.~Dorokhov\thanksref{CORR}},
\author[Zurich]{C.~Amsler},
\author[Purdue]{D.~Bortoletto},
\author[Zurich]{V.~Chiochia},
\author[Miss]{L.~Cremaldi},
\author[Basel]{S.~Cucciarelli},
\author[Basel]{M.~Konecki},
\author[Zurich,PSI]{K.~Prokofiev},
\author[Zurich]{C.~Regenfus},
\author[PSI]{T.~Rohe},
\author[Miss]{D.~Sanders},
\author[Purdue]{S.~Son},
\author[Zurich]{T.~Speer},
\author[JHU]{M.~Swartz}
\address[Zurich]{Physik Institut der Universit\"at Z\"urich-Irchel, 8057 Z\"urich, Switzerland}
\address[PSI]{Paul Scherrer Institut, 5232 Villingen PSI, Switzerland}
\address[Purdue]{Purdue University, Task G, West Lafayette, IN 47907, USA}
\address[Miss]{Mississippi State Univ., Department of Physics and Astronomy, MS 39762, USA}
\address[Basel]{Institut f\"ur Physik der Universit\"at Basel, Basel, Switzerland}
\address[JHU]{Johns Hopkins University, Baltimore, MD, USA}
\thanks[CORR]{Corresponding author. {\it E-mail address:} Andrei.Dorokhov@cern.ch (A.Dorokhov).}

\begin{abstract}
The tracking system of the CMS experiment, currently under construction at the Large
Hadron Collider (LHC) at CERN (Geneva, Switzerland), will include a silicon pixel detector
providing three spacial measurements in its final configuration  for tracks produced 
in high energy $pp$ collisions.
In this paper we present the results of test beam measurements performed at CERN
on irradiated silicon pixel sensors.
Lorentz angle and charge collection efficiency 
were measured for two sensor designs and at various bias voltages.
\end{abstract}
\begin{keyword}
Pixel; Radiation hardness; Lorentz angle; Charge collection; CMS; LHC
\end{keyword}
\end{frontmatter}

\section{Introduction}
The CMS experiment at the Large Hadron Collider (LHC) at CERN (Geneva, Switzerland) 
will be equip\-ped with a silicon pixel detector providing high resolution 3D coordinates 
of the tracks. It will consist of three layers of pixel sensors 
(100~\um~$\times $ 150~\um~pitch size) 
in the barrel region and two discs in the forward regions~\cite{CMS:TDR}.  
The innermost barrel layer is expected to be
exposed to a fluence\footnote{All particle fluences are normalized to 
1 MeV neutrons (${\rm n_{eq}}/\mbox{cm}^2$).} of ~$3 \times 10^{14}~{\rm n_{eq}}/\mbox{cm}^2$ per year
at full luminosity. The second and third layer will be exposed to ~$1.4 \times 10^{14}~{\rm n_{eq}}/\mbox{cm}^2$
and $0.6 \times 10^{14}~{\rm n_{eq}}/\mbox{cm}^2$ per year, respectively. 
The charge sharing induced by the Lorentz force in the 4~T CMS detector magnetic field 
can improve the spatial resolution.
Hence it is very important to study the effects of charge collection efficiency
and Lorentz drift in irradiated silicon devices. 
Here we present measurements of Lorentz angle and charge collection efficiency,
and compare two sensor designs at different irradiation doses.

\section{Experimental setup}
\begin{figure}[htb]
  \begin{center}
    \epsfig{file=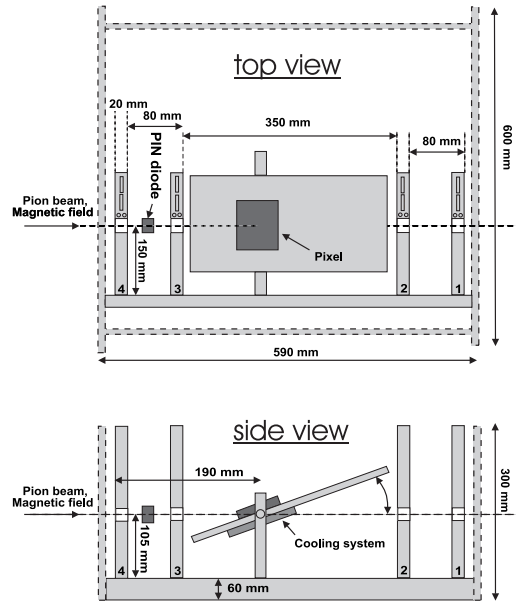,width=1.07\linewidth,clip=,angle=0,silent=}
  \end{center}
  \caption{Top and side views of the beam telescope.}
    \label{FIG:beam_telescope}
\end{figure}

The measurements were performed in the H2 beam line of the 
CERN SPS in June and September 2003  using 150-225 GeV pions. A silicon reference 
telescope~\cite{Amsler:ta} was used to allow a precise
determination of the particle hit position in the pixel detector.
Both, the telescope modules and the pixel front-end, were mounted onto a 
common frame (see fig.~\ref{FIG:beam_telescope}).
The beam telescope consists of 4 modules.
Each consists of two 300~\um~thick single-sided silicon detectors ($32 \times 30$ mm$^2$)
with a strip pitch of 25~\um~(readout pitch of 50~\um), which are oriented perpendicularly.
The resulting intrinsic resolution of the beam telescope is around 1~$\upmu$m. \\
The pixel sensor with the readout chip is mounted on a rotating support positioned between the 
second and the third module. A trigger signal is generated by a silicon PIN diode. 
The data acquisition system, including slow control (temperature and bias voltage of 
the pixel sensor) is written in LabView and LabWindows CVI (National Instruments) running on a PC. 
The analog signals are digitized in a VME based readout system by two CAEN (V550)
and  one custom built FADCs~\cite{Pernicka:pixel_readout}.
The whole setup was placed in a open 3~T Helmholtz magnet with magnetic field parallel 
to the beam. The pixel sensors were cooled down up to $-30^\circ$~C 
by means of water cooled Peltier elements.

\section{Pixel sensors and front-end electronics}
\begin{figure}[htb]
  \begin{center}
    \epsfig{file=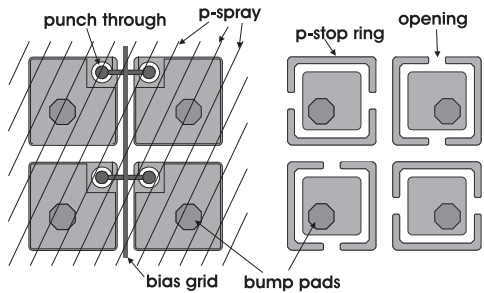,width=\linewidth,clip=,angle=0,silent=}
  \end{center}
  \caption{Layout of the pixel sensors under study: p-spray with bias grid ({\it left}) 
    and open p-stop rings ({\it right}).}
    \label{FIG:designs}
\end{figure}

The CMS pixel sensors are manufactured in the 
"n-on-n" technique, consisting of $ {\rm n}^+$ structures on n bulk silicon.
This allows a partially depleted operation of highly irradiated 
sensors after type inversion, but requires inter-pixel isolation. 
Two isolation techniques were considered in our 
latest prototype designs: p-spray~\cite{Richter:fs}, where a uniform 
medium dose of 
p-impurities covers the whole structured surface, and p-stop, 
where higher dose 
rings individually surround the ${\rm n}^+$-implants. 
Fig.~\ref{FIG:designs} shows the layouts of these designs.
Due to possible failures in the pixel contacts (bump-bonding) 
an intrinsic biasing method has to be introduced.
This is realized by a bias grid and punch-trough structures \cite{Rohe:ww} for the
p-spray detectors and by openings in the rings for the p-stop detectors.
All our test devices had $22\times 32$ pixels, with a sensitive
area of $2.75 \times 4 $ mm$^2$, and a thickness of 280~\um. The 
readout pitch was 125~\um~$\times~$125~\um. \\
Some of the devices were irradiated in a 24 GeV proton beam at the
CERN PS. The sensors received total particle fluences of
$3.3 \times 10^{14}~{\rm n_{eq}}/\mbox{cm}^2$, $8.1 \times 10^{14}~{\rm n_{eq}}/\mbox{cm}^2$
and $1.1 \times 10^{15}~{\rm n_{eq}}/\mbox{cm}^2$ respectively. During the irradiation
the sensors were kept at room temperature and without bias, while after irradiation 
we stored and operated them at $-20^\circ$~C. 
$IV$-curves were measured~\cite{Rohe:IEEE} before and after irradiation,
all measured sensors showed a breakdown voltage above 600 V. 
Finally the sensors were bump bonded to non irradiated 
readout chips of the type {\tt PSI30/AC30}~\cite{Meer:psi30} allowing for a 
non zero-suppressed analog readout of all 704 pixel cells. 
A fast ($20$~ns) external signal for the internal sample and hold mechanism had to be
provided from the trigger PIN diode.

\section{Beam test measurements}

\subsection{Lorentz angle measurements}
\begin{figure}[htb]
  \begin{center}
    \epsfig{file=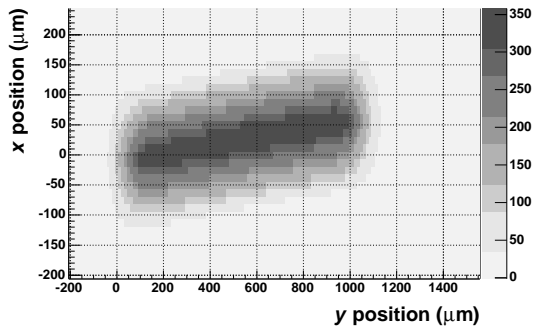,width=1.1\linewidth,clip=,angle=0,silent=}
  \end{center}
  \caption{Deflection of the collected charge in a 3T magnetic field.}
    \label{FIG:la_cluster}
\end{figure}

\begin{figure}[htb]
  \begin{center}
    \epsfig{file=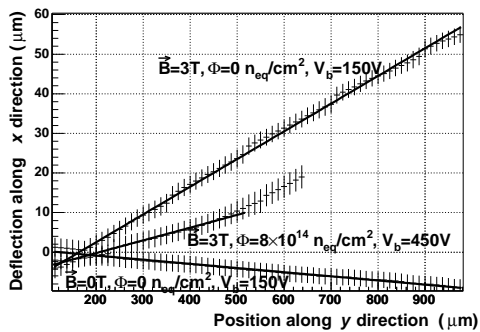,width=1.1\linewidth,clip=,angle=0,silent=}
  \end{center}
  \caption{Deflection of the collected charge as a function of the $y$-position(solid line is fit).}
    \label{FIG:deflection_cluster}
\end{figure}

\begin{figure}[htb]
  \begin{center}
    \epsfig{file=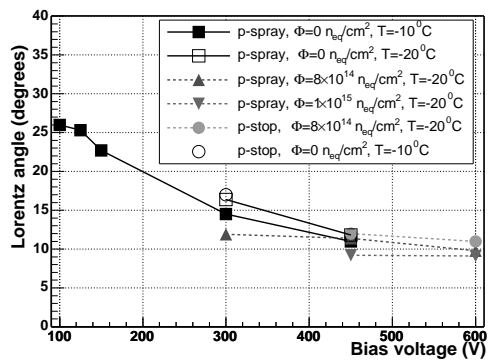,width=1.1\linewidth,clip=,angle=0,silent=}
  \end{center}
  \caption{Lorentz angle as a function of the bias voltage for 4~T magnetic field.}
    \label{FIG:la}
\end{figure}

The Lorentz angle is obtained by the direct measurement of the charge drift in the magnetic field
using the grazing angle method~\cite{Henrich:grazing_angle}.
\begin{figure}[htb]
  \begin{center}
    \epsfig{file=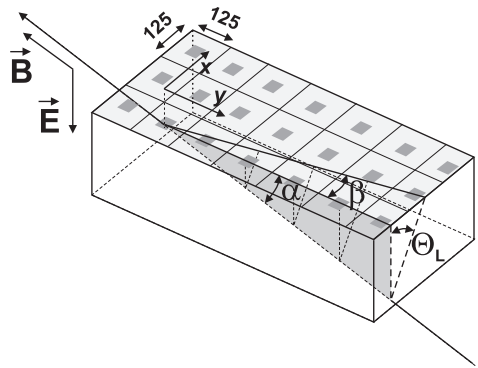,width=\linewidth,clip=,angle=0,silent=}
  \end{center}
  \caption{Lorentz angle measurement with the grazing angle method.}
    \label{FIG:gr_angle}
\end{figure}
The particle beam (parallel to the magnetic field) hits the pixel surface 
at a shallow angle $\alpha=15^\circ$ (see fig.~\ref{FIG:gr_angle}).
The deposited charge drifts according to the combined electric and magnetic forces, 
resulting in a deflection of the particle track projection on the surface by the angle $\beta$.
The Lorentz angle is obtained from the relation $\tan \Theta_L ={\tan \beta / \tan \alpha}$.\\
The deflection measured at 3~T magnetic field is shown in fig.~\ref{FIG:la_cluster} 
for the non-irradiated p-spray sensor. 
The angle $\beta$ is measured by slicing the histogram perpendicularly to the $y$-axis.
The position of the center of each slice is shown in fig.~\ref{FIG:deflection_cluster}
as a function of the position along the $y$-axis. A measurement without magnetic field 
is used to correct for the detector misalignment with respect to the 
beam (bottom line in fig.~\ref{FIG:deflection_cluster}).
For the irradiated device in fig.~\ref{FIG:deflection_cluster} there are
two regions with a different slope, and hence, two different values of the Lorentz angle. 
The difference in the Lorentz angle can be explained by nonuniform 
distribution of the electric field in the irradiated devices. 
Different models of the electric field distribution in the irradiated silicon devices 
are discussed in Refs.~\cite{Eremin:two_peak,Castaldini:double_junction}.
The evidence of the non-uniform electric filed will be shown also in the next section. \\   
As most of the signal charge is collected from the region close to the pixel implant,
this region was used to determine the Lorentz angle.
Figure~\ref{FIG:la} shows the measured values for both non-irradiated and irradiated sensors
extrapolated to 4~T magnetic field.   
We observe a strong dependence of the Lorentz angle on the bias voltage (electric field), 
while it is only weakly affected by the irradiation or sensor design.
For the non-irradiated sensors a Lorentz angle of $26^\circ$ can be reached at a bias voltage of $100$ V, 
while irradiated sensors have to be operated at higher bias voltages, where the Lorentz angle
drops to roughly $10^\circ$. 
Since the electron mobility increases with decreasing temperature the Lorentz
angle measured at the lower temperature $-20^\circ$~C is $1^\circ-2^\circ$ larger.
Our values for the Lorentz angle are in a good agreement with 
measurements and simulations in Refs.~\cite{Bartsch:la1,Bartsch:la2}.
\subsection{Charge collection in irradiated sensors}
\begin{figure}[htb]
  \begin{center}
    \epsfig{file=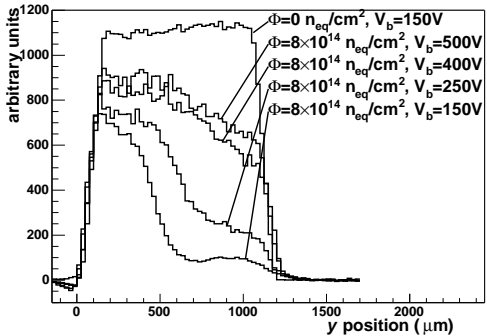 ,width=1.1\linewidth,clip=,angle=0,silent=}
  \end{center}
  \caption{Charge cluster for different bias voltages.}
    \label{FIG:proj_cluster}
\end{figure}

\begin{figure}[htb]
  \begin{center}
    \epsfig{file=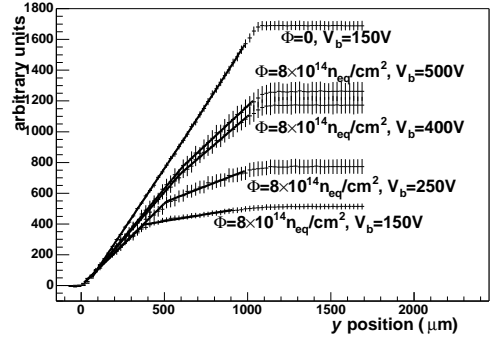,width=1.1\linewidth,clip=,angle=0,silent=}
  \end{center}
  \caption{Integrated charge as a function of the distance to the particle entry point.}
    \label{FIG:int_cluster}
\end{figure}

\begin{figure}[htb]
  \begin{center}
    \epsfig{file=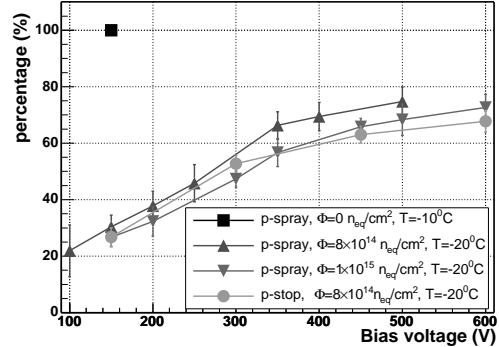,width=1.1\linewidth,clip=,angle=0,silent=}
  \end{center}
  \caption{Total collected charge normalized to non-irradiated device.}
    \label{FIG:q}
\end{figure}

After irradiation the collected charge decreases due to charge trapping 
and partial depletion of the sensor.
The measurements of the charge collection efficiency as a function of the sensor depth
were performed using the grazing angle method without magnetic field.
The averaged charge cluster profiles in non-irradiated and irradiated p-spray 
sensors are shown in fig.~\ref{FIG:proj_cluster}.
In the unirradiated sensor the charge is collected uniformly across the 
whole sensor depth, while the irradiated devices shows two regions with different 
collected charge. 
At low bias voltage in the irradiated detectors 
some charge is also collected from the side opposite to
the pixel implants. This can indicate that the depletion starts from both sides
of the detector, because of the nonuniform distribution of the electric filed.  
The integral of the collected charge from the particle entry point
up to a point along the $y$-coordinate is shown in fig.~\ref{FIG:int_cluster}.
There are two regions with different charge collection, 
represented by two different slopes of the curve for the irradiated device.
The total collected charge for different designs and irradiation doses as a function of bias voltage
is shown in fig.~\ref{FIG:q}.
\subsection{Signal-to-noise ratio}
\begin{figure}
  \begin{center}
    \begin{minipage}[t]{0.5\linewidth}
      \epsfig{file=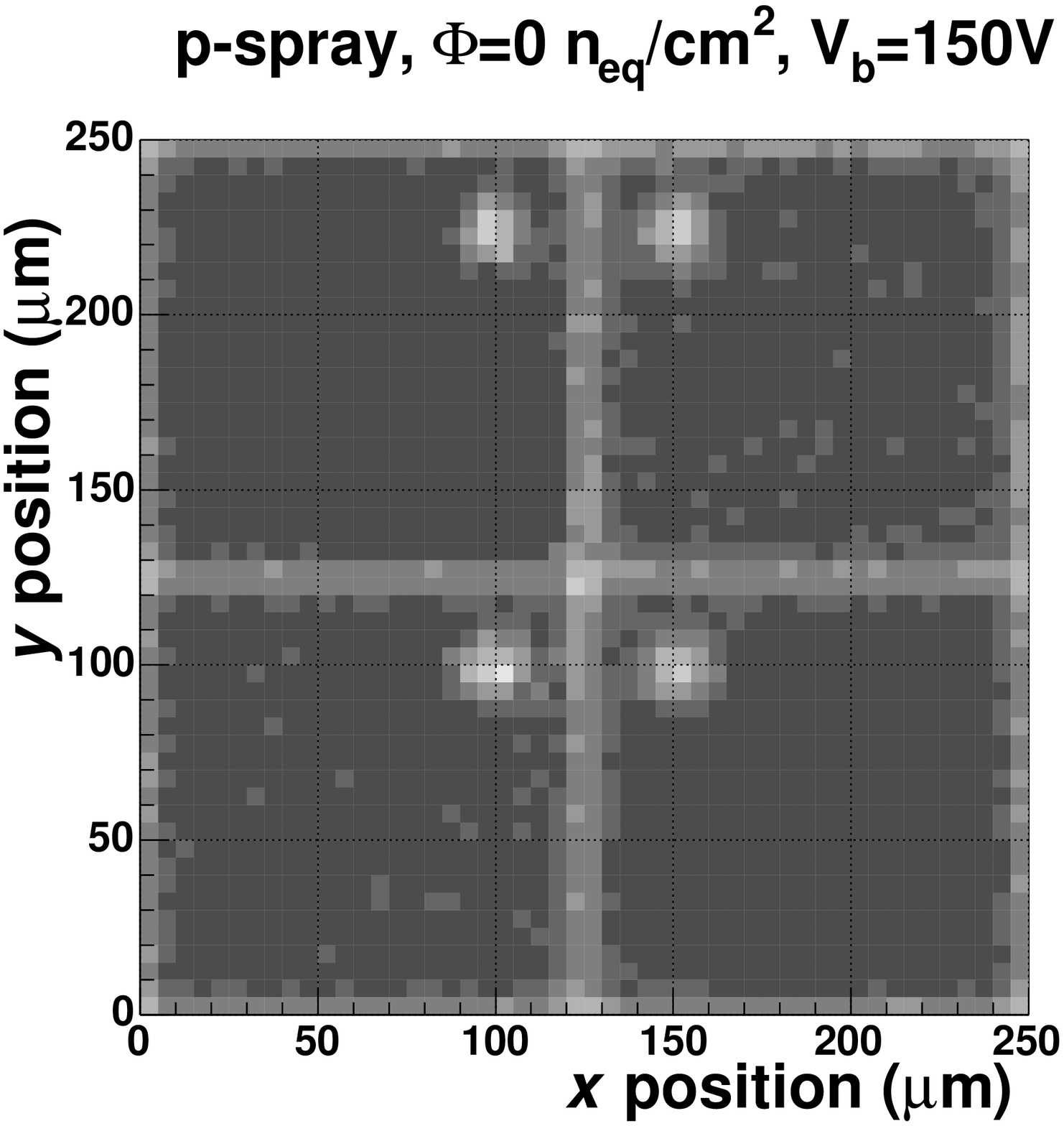, width=1.1\linewidth}
    \end{minipage}\hfill
    \begin{minipage}[t]{0.5\linewidth}
      \epsfig{file=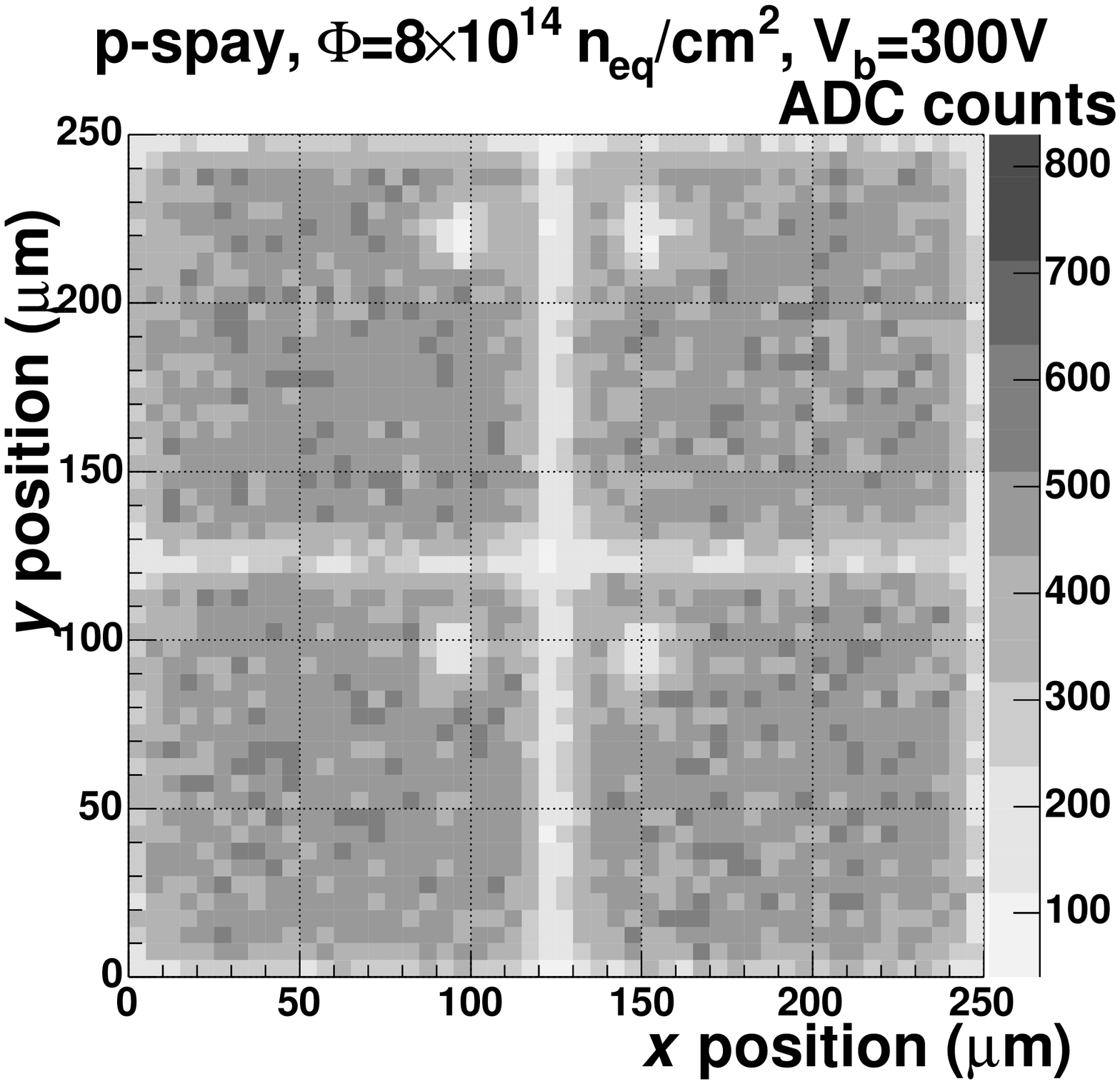, width=1.1\linewidth}
    \end{minipage}

    \begin{minipage}[t]{0.5\linewidth}
      \epsfig{file=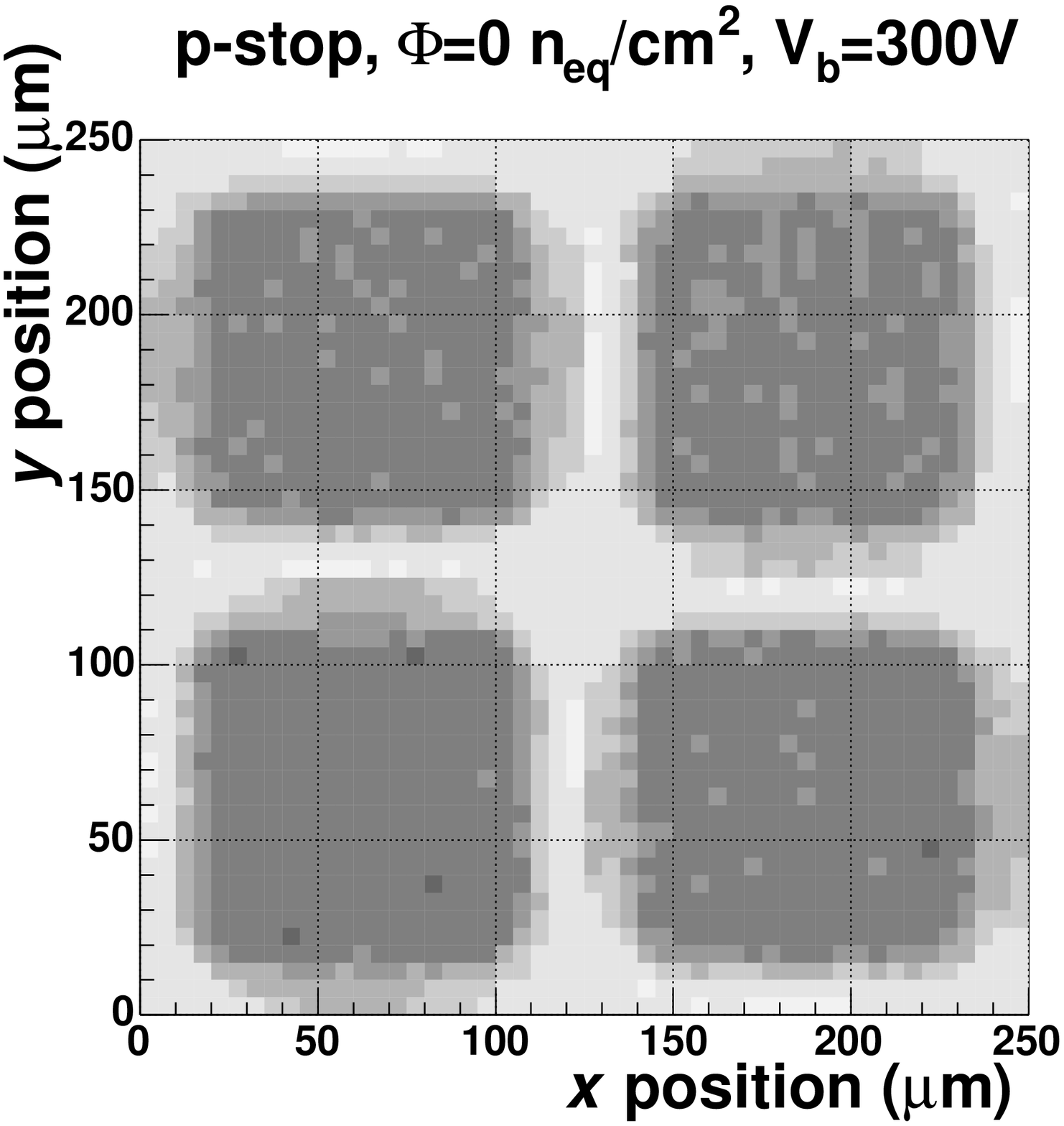, width=1.1\linewidth}
    \end{minipage}\hfill
    \begin{minipage}[t]{0.5\linewidth}
      \epsfig{file=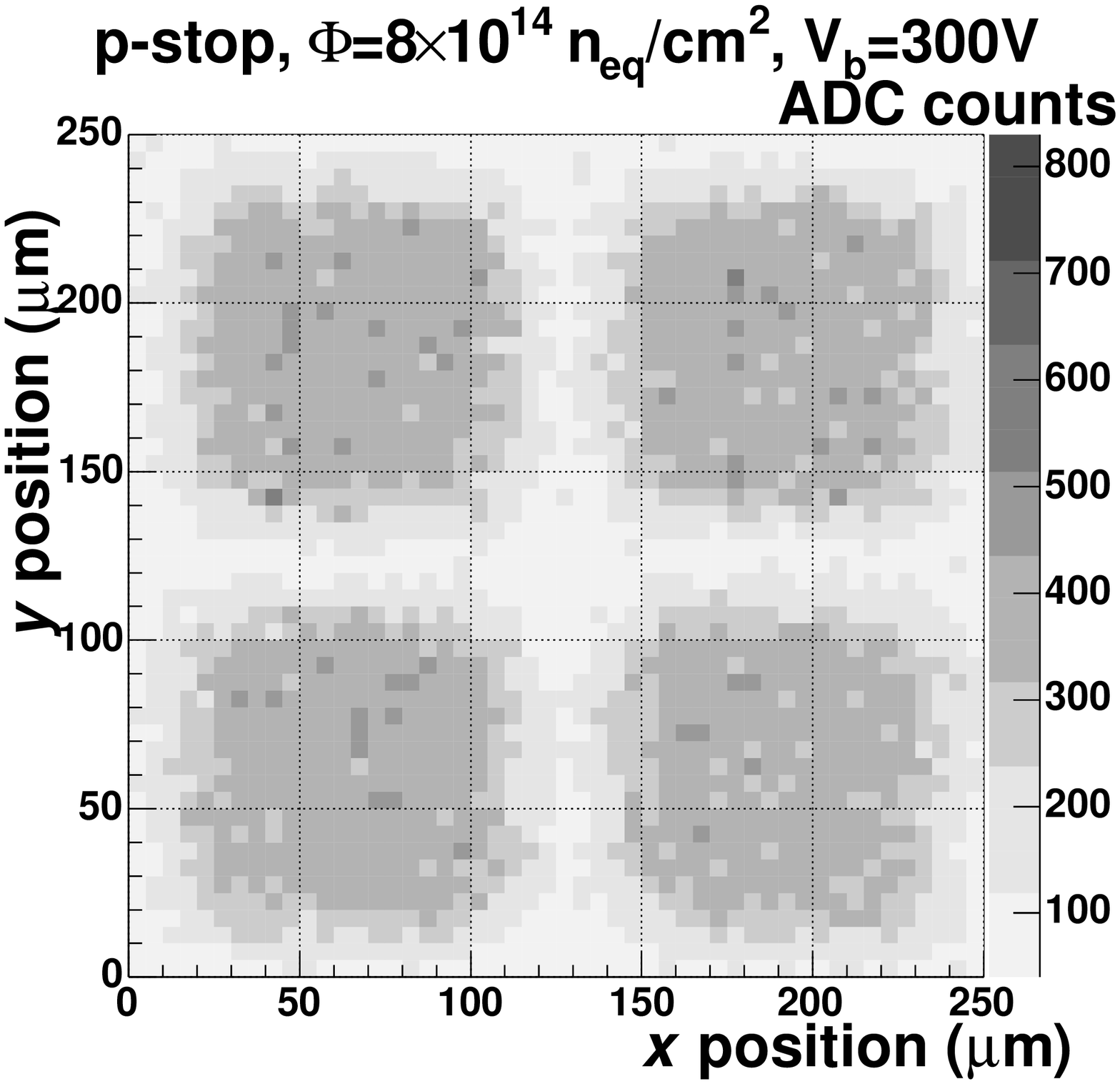, width=1.1\linewidth}
    \end{minipage}
  \caption{Charge collected in the hit pixel for non irradiated (left) 
and irradiated sensors (right).}
    \label{FIG:charge_dist}
  \end{center}
\end{figure}

\begin{figure}[htb]
  \begin{center}
    \epsfig{file=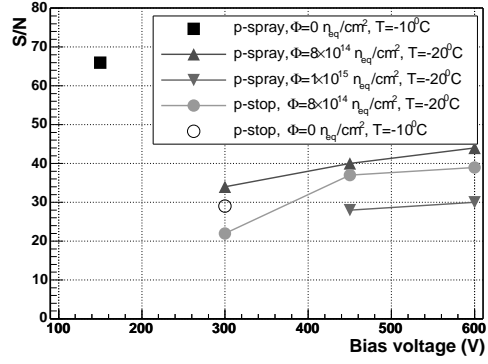,width=1.1\linewidth,clip=,angle=0,silent=}
  \end{center}
  \caption{Signal-to-noise ratio as a function of the bias voltage.}
    \label{FIG:sn}
\end{figure}

The charge collection depends on the position of the incident particle
with respect to the pixel implant, caused by the regions with reduced
sensitivity.
The average charge collected in the hit pixel as a function of the hit
position is shown in fig.~\ref{FIG:charge_dist} for non-irradiated
and irradiated sensors.  
One can see that after irradiation the average pixel signal decreases, 
and the area with the reduced charge collection increases
in the case of the p-stop design.\\ 
The signal-to-noise ratio (S/N) averaged over the whole pixel area is shown in fig.~\ref{FIG:sn}. 
The measured signal-to-noise ratio is above 30 even after the irradiation.
We observe lower S/N ratio in the unirradiated p-stop device,
because of the charge spread over several pixels caused by the p-stop openings.
A more detailed study of charge collection for these sensors 
can be found in Ref.~\cite{Rohe:IEEE}.   
\section{Conclusions}
Two different designs of the CMS prototype pixels sensors were tested 
up to 600~V bias voltage and after exposure to particle fluences up to
$1.1 \times 10^{15}~{\rm n_{eq}}/\mbox{cm}^2$.
For the irradiated devices we observe two regions across the sensor
with different charge collection and value of the Lorentz angle.
This result is an indication of a nonuniform electric field in the irradiated devices.\\
The Lorentz angle mainly depends on the bias voltage
and it shows a weak dependence on the irradiation dose or sensor design.
The Lorentz angle at 4 T magnetic field reaches 26$^\circ$
for the non-irradiated devices at a bias voltage 
of 100~V and it drops to 8.3$^\circ$ for the ones irradiated 
at $1.1 \times 10^{15}~{\rm n_{eq}}/\mbox{cm}^2$ and at a bias voltage 600~V.\\
After irradiation and at high bias voltage (600 V)
the charge collection is about 60\% of the value observed for the non-irradiated devices,
and it is slightly larger for the p-spray design.\\
The signal-to-noise ratio decreases from 65 to about 35 after irradiation
giving us confidence in operating the CMS
pixel detector up to the maximum expected irradiation dose.\\ 
For the irradiated devices the charge collection, and therefore the signal height, 
increases with the bias voltage but the Lorentz angle decreases.
For this reason the bias voltage must be optimized 
for the best performance of the CMS pixel detector.\\

\section*{Acknowledgments}
We gratefully acknowledge Silvan Streuli from ETH Zurich and Fredy Glaus from PSI for
their immense effort on the bump bonding, Maurice Glaser and Michael Moll from CERN for 
carrying out the irradiation, Kurt B\"osiger from the Z\"urich workshop for the mechanical
construction, Gy\"orgy Bencze and Pascal Petiot from CERN for the H2 beam line support 
and finally the whole CERN-SPS team. The authors would like to thank Dr.~Danek Kotlinski from PSI for very
useful discussions.

\end{document}